# A method to generate complex quasi-nondiffracting optical lattices

Servando López-Aguayo,[1,2] Yaroslav V. Kartashov,[1] Victor A. Vysloukh,[3] and Lluis Torner[1]

[1]*ICFO-Institut de Ciencies Fotoniques, and Universitat Politecnica de Catalunya, Mediterranean Technology Park, 08860 Castelldefels (Barcelona), Spain*

[2]*Photonics and Mathematical Optics Group, Tecnólogico de Monterrey, Monterrey México 64849*

[3]*Departamento de Fisica y Matematicas, Universidad de las Americas – Puebla, 72820, Puebla, Mexico*

We put forward a powerful technique that allows generating quasi-non-diffracting light beams with a variety of complex transverse shapes and topologies. We show that, e.g., spiraling patterns, patterns featuring curved or bent bright stripes, or patterns featuring arbitrary combinations of harmonic, Bessel, Mathieu and parabolic beams occupying different domains in the transverse plane can be produced. The quasi-non-diffracting patterns open up a wealth of opportunities for the manipulation of matter and optical waves, colloidal and living particles, with applications in biophysics, and quantum, nonlinear and atom optics.

*PACS numbers: 42.65.Jx; 42.65.Tg; 42.65.Wi*

The advent of optical trapping and manipulation of matter has revolutionized several branches of physics from the micro- and nano-scale to the single-atom levels and Bose-Einstein condensates [1]. Nondiffracting light patterns have become key tools in topics as diverse as trapping of in-vivo and colloidal particles in biophysics [2], atom optics [3], applications of optical lattices for quantum computing [4] and quantum optics at large [5], optical tweezing [6], and nonlinear optics [7,8], to name a few. The patterns used to date correspond only to the known sets of simple non-diffracting light beams that are rigorous exact solutions of Helmholtz equation. In particular, group theory demonstrates that there are only four different coordinate systems where Helmholtz equation is separable [9], yielding invariant solutions along the propagation axis: plane waves in Cartesian coordinates, Bessel beams in circular cylindrical coordinates [10], Mathieu beams in elliptic cylindrical coordinates [11], and parabolic beams in parabolic cylindrical coordinates [12]. In addition one can also mention accelerating Airy beams [13]. Each of these beams exhibits a specific symmetry, hence affording the corresponding possibilities and limitations. An important related open problem is the generation of more complex nondiffracting, or slowly diffracting, beams with arbitrary shapes and symmetries. Here we put forward a powerful new strategy that allows the generation of arbitrary complex light patterns matching the requirements of a particular application, which can be considered nondiffracting for all practical purposes. Such complex beams diffract very slowly, so that they can be considered as non-diffracting over a distance dictated by the width of the angular spectrum.

The field of a general non-diffracting beam propagating along the $\xi$ axis that does not experience acceleration in the transverse plane may be written via the Whittaker integral [10-12]:

$$q_{\text{latt}}(\eta,\zeta,\xi) = \exp(-ik_\xi \xi)\int_0^{2\pi} G(\varphi)\exp[ik_{\text{t}}(\eta\cos\varphi + \zeta\sin\varphi)]d\varphi. \qquad (1)$$

Here $k_\xi$ and $k_{\text{t}}$ are longitudinal and transverse components of the wave number $k = (k_\xi^2 + k_{\text{t}}^2)^{1/2}$, respectively, $\varphi$ is the azimuthal angle in frequency space, $\eta, \zeta$ are the transverse coordinates and $G(\varphi)$ is the angular spectrum which is defined on an infinitely narrow ring of radius $k_{\text{t}}$. In experiments, truncated versions of nondiffracting beams are commonly used that still can be considered nondiffracting up to a finite distance. If the nondiffracting beam is modulated by a Gaussian envelope, such distance is $\sim w_0 k / k_{\text{t}}$, where $w_0$ is the radius of the envelope. Such beams have an angular spectrum defined on an annular ring of radius $k_{\text{t}}$ with width $\sim 4/w_0$ [14]. A finite width of the angular spectrum does not necessarily imply truncation of the pattern. Superposition of two infinitely extended Bessel beams with slightly different $k_{\text{t}}$ generates a pattern that can be considered undistorted over a distance $\xi$ that is dictated by the difference in the $k_{\text{t}}$ values. Such a pattern will experience distortion in the entire transverse plane due to the accumulated phase difference between the fields, in contrast to truncated patterns where the perturbation moves from the periphery to beam center. The point is increasing the width of the angular spectrum in frequency space opens up the possibility to construct beams with really complex shapes.

Our approach consists in engineering the angular spectrum in the frequency space under the constraint that the transverse wavenumber components $k_\eta, k_\zeta$ ($k_{\text{t}}^2 = k_\eta^2 + k_\zeta^2$) are contained within a sufficiently narrow annular ring to ensure almost nondiffracting propaga-tion. The experimental feasibility of such a concept has been demonstrated [15]. Here we put forward an iterative Fourier algorithm for construction of beams with arbitrarily complex shapes that is reminiscent to methods used in phase retrieval and image processing algorithms [16]. The first step is setting the desired field distribution $\tilde{q}(\eta,\zeta)$ at $\xi=0$. The phase distribution $\arg[\tilde{q}(\eta,\zeta)]$ of the field is a free parameter, while $|\tilde{q}(\eta,\zeta)|$ is selected to get the desired shape. Quasi-random (or uniform) initial phase distributions yield convergence in most cases. However, an initial guess intuitively adapted to the desired final $\arg[q(\eta,\zeta)]$ distribution accelerates convergence. On the next step the Fourier transform of $\tilde{q}(\eta,\zeta)$ is calculated and the components of the angular spectrum for $k_\eta, k_\zeta$ falling outside the annular ring of width $\delta k_\text{t}$ and radius $k_\text{t}$ are set to zero. One applies an inverse Fourier transform to the resulting function and substitutes the modulus of the obtained complex function with the original field modulus $|\tilde{q}(\eta,\zeta)|$, but keeps the new phase distribution. This procedure is repeated until convergence is achieved for a selected $\delta k_\text{t}$. The phase factor $\exp(i\varphi_0)$ in the trial distribution $\tilde{q}(\eta,\zeta)$ does not affect convergence. The field $q(\eta,\zeta)$ from the last iteration is used without replacing its modulus with $|\tilde{q}(\eta,\zeta)|$, so that some distortions will always appear in $|q(\eta,\zeta)|$, in comparison with the ideal distribution. The iterative procedure produces patterns involving combinations of multiple truly nondiffracting beams with slightly different $k_\text{t}$ values as dictated by the imposed width $\delta k_\text{t}$ of the angular spectrum. Such an iterative procedure is crucial: the propagated trial beam $\tilde{q}(\eta,\zeta)$ decays after just a few diffraction lengths, while the iterated beam keeps its structure over tens of diffraction lengths. We use dimensionless transverse coordinates $\eta, \zeta$ normalized to the characteristic width $r_0$, while the longitudinal coordinate $\xi$ is normalized to the diffraction length $L_\text{dif} = k_0 r_0^2$, where $k_0 = 2\pi/\lambda$ is the wavenumber. Thus, a beam at the wavelength $\lambda = 532$ nm shaped in accordance with our method that has a characteristic transverse scale [for example, a spacing between stripes in Fig. 2(a)] of the order of $r_0 \sim 10$ $\mu$m will remain undistorted over distance considerably exceeding $L_\text{dif} \sim 1.2$ mm, while for the beam with $r_0 \sim 1$ mm the distance of invariance will exceed $L_\text{dif} \sim 12$ m.

Examples of patterns generated with this algorithm are shown in Fig. 1, where we aim to produce specific spiraling beams. For a very small width of the angular spectrum $\delta \sim 0.01$ ($\delta = \delta k_\text{t}/k_\text{t}$) one usually gets patterns that are far from the desired ones, especially when $\tilde{q}(\eta,\zeta)$ exhibits a complicated structure [Fig. 1(a)]. Increasing $\delta$ up to 0.1 causes dramatic improvements in the beam shape: while some distortions are still visible, the desired spiraling pattern is clearly resolvable [Fig. 1(b)]. Thus, engineering the angular spectrum allows to construct *patterns that have no analogs among known non-diffracting beams.*

If $\delta$ is further increased one obtains even better approximation to the desired beam [Fig. 1(c)]. However, the value of $\delta$ has to be carefully selected since a small $\delta k_t$ assures almost diffractionless propagation, but at the same time it may result in patterns that are rather far from the desired ones, while for sufficiently large $\delta k_t$ one can generate patterns close to any desired beam that, however, will be more prone to diffraction. Still, in previous experiment [14] it was demonstrated that Bessel beams with a Gaussian envelope may propagate undistorted over distances largely exceeding the diffraction length even for $\delta \sim 0.2$.

Our technique allows to introduce controllable distortion into otherwise rigorous non-diffracting beams. Thus, using a trial function $\tilde{q} = \cos(k_t \eta)$ for $\zeta > 0$ and $\tilde{q} = \cos[k_t(\eta \cos\theta_b + \zeta \sin\theta_b)]$ for $\zeta \leq 0$ one can generate a quasi-one-dimensional beam with stripes experiencing an abrupt bending at an angle $\theta_b$ at $\zeta = 0$ [Fig. 2(a)]. Due to inherent robustness of the method the sharp shape variations around $\zeta = 0$ are smoothed out. While for small angles of bending $-\pi/18 \leq \theta_b \leq \pi/18$ the beam shape is remarkably regular and its intensity remains almost unchanged along the stripes, for higher bending angles the regions of increased or decreased intensity appear [Fig. 2(b)]. Deformed patterns featuring stripes that may periodically curve in horizontal direction are produced with $\tilde{q} = \cos[k_t \eta \cos\theta_b + \delta a \cos(k_t \zeta \sin\theta_b)]$, where $\delta a$ controls the amplitude of deformation. For sufficiently small deformations $\delta a$ the resulting quasi-nondiffracting beams feature almost constant intensities along stripes [Fig. 2(c)], while increasing $\delta a$ results in appearance of domains with increased or decreased intensities and the actual bending law for beam stripes may depart from the harmonic one [Fig. 2(d)].

The method allows also the identification of shapes of angular spectra corresponding to novel types of non-diffracting beams. Thus, a trial beam $\tilde{q} = J_0(k_t \eta \cos\theta_b)\exp(ik_t \zeta \sin\theta_b)$, with $J_0$ being zero-order Bessel function, allows to produce a single-channel pattern in real space [Fig. 2(e)], while in frequency domain the spectrum of such beam appears to be very close to infinitely-narrow ring and its angular distribution can be well described by a step-like function $G(\varphi)$ that is nonzero within finite interval of angles $\varphi_1 < \varphi < \varphi_2$. This indicates that there exist truly nondiffracting beams with such specific symmetry. In a similar way one can construct truly nondiffracting beams featuring several pronounced stripes [Fig. 2(f)]. The technique may generate patterns featuring practically any combinations of known harmonic, Bessel, Mathieu, or parabolic beams occupying different arbitrary domains in the transverse plane that propagate undistorted over considerable distances. Thus, the trial beam $\tilde{q} = Je_m(k_t, \varepsilon)ce_m(k_t, \varepsilon) + iJo_m(k_t, \varepsilon)se_m(k_t, \varepsilon)$ for $\phi_1 < \phi < \phi_2$ and $\tilde{q} = 0$ otherwise, where $Je_m, Jo_m$ are even and odd radial Mathieu functions, $ce_m, se_m$ are even and odd angu-lar Mathieu functions, $\varepsilon$ is the ellipticity parameter, and $\phi$ is the azimuthal angle in spatial domain, produces the pattern featuring several confocal elliptical rings in a selected angular domain $\phi_1 < \phi < \phi_2$, while in other angular domain the light field vanishes almost completely [Figs. 3(a) and 3(b)]. The symmetry of the Mathieu pattern for $\phi_1 < \phi < \phi_2$ is almost unaffected by removal of part of the beam in other angular domain. Increasing of the angular spectrum width $\delta$ from 0.1 [Fig. 3(a)] to 0.2 [Fig. 3(b)] results only in slight modifications in the beam. Such states experience exceptionally slow transformations on propagation, i.e. they are very close to nondiffracting beams. The possibility to combine beams with different symmetries is illustrated in Figs. 3(c)-3(f) where a parabolic trial beam $\tilde{q} = \text{Pe}(k_t, a)\text{Pe}(k_t, -a) + i\text{Po}(k_t, a)\text{Po}(k_t, -a)$ defined at $\eta < 0$ (here Pe, Po are the even and odd parabolic cylinder functions, respectively, and parameter $a$ determines the curvature of beam stripes) was combined either with harmonic $\tilde{q} = \psi_b \cos(k_t \eta)$ or Bessel $\tilde{q} = \psi_b \text{J}_1(k_t r)$ patterns at $\eta \geq 0$ (here $r$ is the radial coordinate and $\psi_b$ determines the ratio of beam amplitudes at $\eta < 0$ and $\eta > 0$). The resulting quasi-nondiffracting beams are characterized by sharp transitions between domains with different field symmetries. An example of more complicated almost nondiffracting pattern that does not have analogs among truly nondiffracting beams is presented in Fig. 4. The beam of this type is produced by $\tilde{q} = \sin(k_t r - n\phi)$, where $n = 0,1,2,...$ is an integer. When $\delta = 0.25$ and $n = 0$ a pattern is generated [Fig. 4(a)] whose shape is well described by radially periodic cosine function. For $n = 1,...,5$ and $\delta = 0.25$ the method generates different types of spiraling beams that are distorted in the center, but are remarkably regular at moderate $r$ values [Figs. 4(b)-4(f)].

Our method can be modified in order to generate a required phase distribution in the beam instead of the field modulus. The straightforward modification of the iterative procedure allows to combine patterns characterized by different topological winding numbers (charges) such as $\tilde{q} = \text{J}_m(k_t r)\exp(im\phi)$ at $0 < \phi < \pi$ and $\tilde{q} = \psi_b \text{J}_n(k_t r)\exp(in\phi)$ at $\pi < \phi < 2\pi$. Thus, in the case of $m = 3$, $n = 1$ the pattern is obtained whose intensity remains almost invariant on propagation, while phase accumulation rates in different halves of the pattern differ considerably [Fig. 5(a)]. Using $m = 3$ and $n = -1$ allows to obtain the beam with opposite phase accumulation rates in adjacent half-planes [Fig. 5(b)]. It is also possible to change the phase distribution not in angular, but in radial direction [Fig. 5(c)]. These results can be immediately used to generate suitable optical tweezers and atom traps, as well as to study the transfer of angular momentum to atoms or microparticles.

The beams described here may be used to demonstrate a variety of effects in different areas of science, from quantum optics to physics of matter waves and nonlinear optics.

Among their applications may be the control of evolution of matter-wave or optical solitons in optical lattices produced by the corresponding nondiffracting beams. Due to their unusual symmetry such lattices may allow observation of new types of soliton motion and may substantially enrich the possibilities for all-optical routing of light signals. To illustrate this, we consider the propagation of optical radiation in a biased photorefractive crystal. The lattice that is optically induced by suitable quasi-nondiffracting beam creates refractive index modulation in the transverse plane $(\eta,\zeta)$ that can be considered invariable in $\xi$ direction for sufficiently small $\delta$ values. Nonlinearity of the crystal affects only the probe beam with polarization orthogonal to that of lattice-creating beam that propagates in linear regime [7]. The propagation of probe beam is described by the normalized nonlinear Schrödinger equation $iq_\xi + (1/2)(q_{\eta\eta} + q_{\zeta\zeta}) + Eq(1 + S|q|^2 + R)^{-1}(S|q|^2 + R) = 0$, where $S = 0.2$ is the saturation parameter, $E = 12$ is the biasing field applied to the crystal and the function $R$ describes the lattice shape that is proportional to intensity of lattice-creating beam. If the optical lattice features clearly pronounced guiding channels in the transverse plane the soliton launched into one of such channels with a proper input phase tilt may start moving along the guiding channel, so that the trajectory of soliton motion will be dictated by the topology of the lattice. In this way one can force solitons to change their propagation direction in lattices with bent channels [Fig. 6(a)], to move along curved trajectory [Fig. 6(c)], or perform specific spiraling motion in spiraling lattices [Fig. 6(c)]. Such dynamics usually is not accessible in conventional truly nondiffracting lattices.

Summarizing, we put forward a technique to generate new types of complex quasi-nondiffracting light patterns. The key ingredient of the method is engineering the angular spectrum of the kernel-generated function. The wider the rings of the angular spectrum the higher the complexity of the patterns generated, but the shorter the propagation distance where they remain undistorted. The light patterns described here are expected to find important applications in several branches of science that currently use non-diffracting light beams for the manipulation of matter, such us optical traps in biophysics and quantum and atom optics, or to manipulate light itself.

# Figure captions

Figure 1. Spatial intensity distributions of spiraling beams (top row) and corresponding angular spectra in frequency space (bottom row) for (a) $\delta = 0.01$, (b) $0.07$, and (c) $0.20$.

Figure 2. Bent beams corresponding to (a) $\theta_b = 0.122$ and (b) $\theta_b = -0.209$ at $k_t = 2$. Curved beams corresponding to (c) $\delta a = -1$ and (d) $\delta a = 2$ at $\theta_b = 0.209$, $k_t = 2$. Quasi-one-dimensional beams with one (e) and three (f) enhanced channels at $k_t = 4$ and $\delta = 0.1$.

Figure 3. Intensity distributions for (a),(b) truncated Mathieu beams, (c),(d) parabolic-cosine beams, and (e),(f) parabolic-Bessel beams. Top panels correspond to $\delta = 0.1$, while bottom panels correspond to $\delta = 0.2$. In all cases $k_t = 4$.

Figure 4. Intensity distributions of spiraling beams at (a) $n = 0$, (b) $n = 1$, (c) $n = 2$, (d) $n = 3$, (e) $n = 4$, and (f) $n = 5$. In all cases $k_t = 4$ and $\delta = 0.25$.

Figure 5. Intensity distributions and engineered phase structures of quasi-nondiffracting beams obtained by using as a trial pattern a combination of two Bessel beams with topological charges (a) $m = +3$, $n = +1$, (b) $m = +3$, $n = -1$, and (c) $m = -5$, $n = -1$. In all cases $k_t = 4$.

Figure 6. Snapshot images showing dynamics of soliton propagation in (a) bent lattices with $\theta_b = \pm 0.087$, $k_t = 2$, (b) in curved lattice with $\delta a = -0.5$, $\theta_b = 0.209$, $k_t = 2$, and (c) in spiraling lattice with $n = 1$, $k_t = 4$. In (a) the intensity distributions corresponding to two different lattices are superimposed.

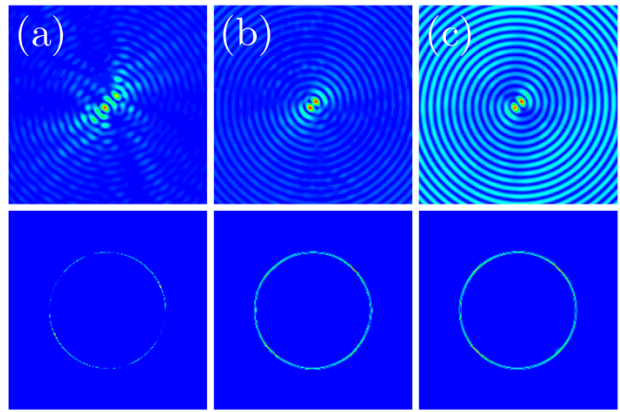

Figure 1. Spatial intensity distributions of spiraling beams (top row) and corresponding angular spectra in frequency space (bottom row) for (a) $\delta = 0.01$, (b) $0.07$, and (c) $0.20$.

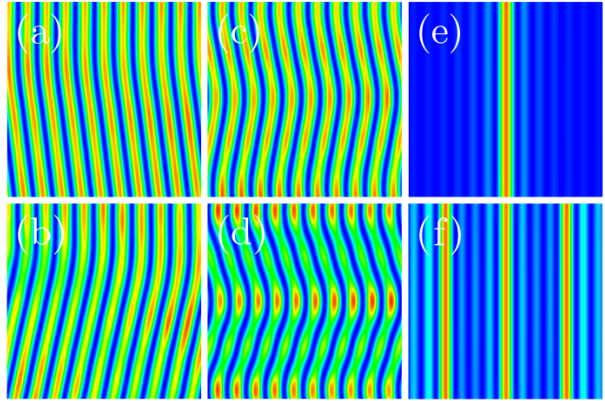

Figure 2.   Bent beams corresponding to (a) $\theta_{\rm b} = 0.122$ and (b) $\theta_{\rm b} = -0.209$ at $k_{\rm t} = 2$. Curved beams corresponding to (c) $\delta a = -1$ and (d) $\delta a = 2$ at $\theta_{\rm b} = 0.209$, $k_{\rm t} = 2$. Quasi-one-dimensional beams with one (e) and three (f) enhanced channels at $k_{\rm t} = 4$ and $\delta = 0.1$.

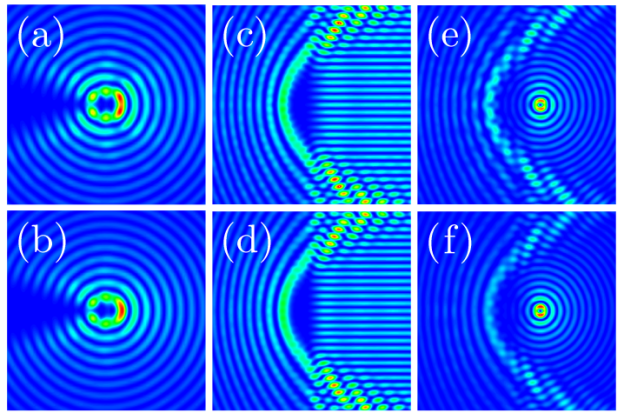

Figure 3. Intensity distributions for (a),(b) truncated Mathieu beams, (c),(d) parabolic-cosine beams, and (e),(f) parabolic-Bessel beams. Top panels correspond to $\delta = 0.1$, while bottom panels correspond to $\delta = 0.2$. In all cases $k_t = 4$.

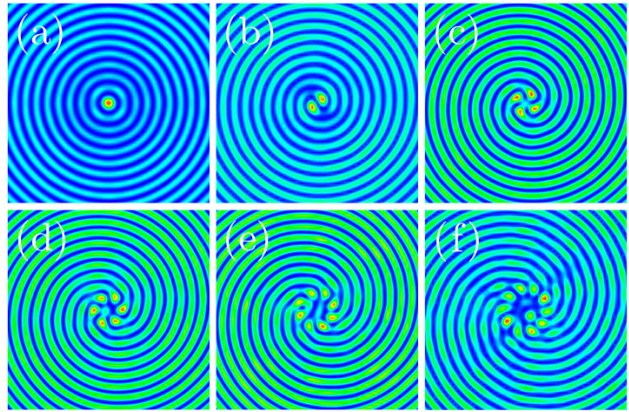

Figure 4. Intensity distributions of spiraling beams at (a) $n=0$, (b) $n=1$, (c) $n=2$, (d) $n=3$, (e) $n=4$, and (f) $n=5$. In all cases $k_t=4$ and $\delta=0.25$.

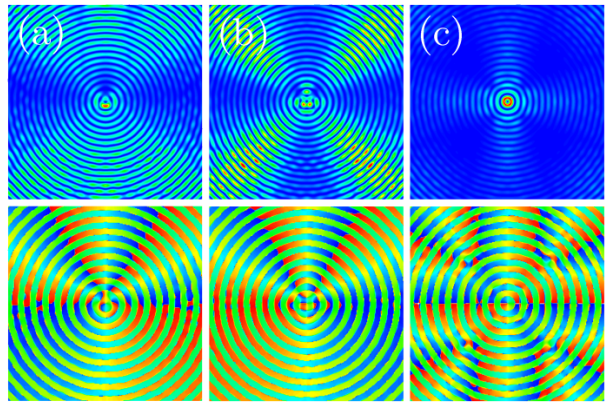

Figure 5. Intensity distributions and engineered phase structures of quasi-nondiffracting beams obtained by using as a trial pattern a combination of two Bessel beams with topological charges (a) $m = +3$, $n = +1$, (b) $m = +3$, $n = -1$, and (c) $m = -5$, $n = -1$. In all cases $k_t = 4$.

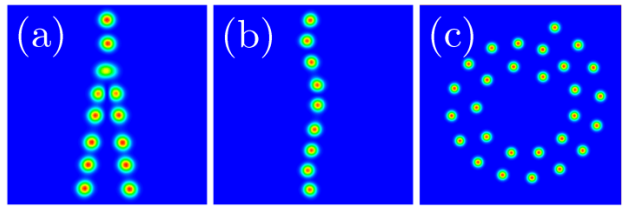

Figure 6. Snapshot images showing dynamics of soliton propagation in (a) bent lattices with $\theta_\mathrm{b} = \pm 0.087$, $k_\mathrm{t} = 2$, (b) in curved lattice with $\delta a = -0.5$, $\theta_\mathrm{b} = 0.209$, $k_\mathrm{t} = 2$, and (c) in spiraling lattice with $n = 1$, $k_\mathrm{t} = 4$. In (a) the intensity distributions corresponding to two different lattices are superimposed.

14